\documentclass[letterpaper,twocolumn,prl,aps,superscriptaddress,amsmath,amssymb,floatfix]{revtex4}
\usepackage{mathptmx}
\usepackage[latin9]{inputenc}
\usepackage{color}
\usepackage{verbatim}
\usepackage{float}
\usepackage{amsmath}
\usepackage{amssymb}
\usepackage{graphicx}
\usepackage{esint}

\usepackage[unicode=true,
 bookmarks=true,bookmarksnumbered=false,bookmarksopen=false,
 breaklinks=false,pdfborder={0 0 1},backref=false,colorlinks=true]
 {hyperref}
\hypersetup{
 linkcolor=magenta,urlcolor=blue,citecolor=blue,pdfstartview={FitH},hyperfootnotes=false}

\makeatletter

\usepackage{textcomp}
\usepackage{epstopdf}

\pdfpageheight\paperheight
\pdfpagewidth\paperwidth

\@ifundefined{textcolor}{}{%
 \definecolor{BLACK}{gray}{0}
 \definecolor{WHITE}{gray}{1}
 \definecolor{RED}{rgb}{1,0,0}
 \definecolor{GREEN}{rgb}{0,1,0}
 \definecolor{BLUE}{rgb}{0,0,1}
 \definecolor{CYAN}{cmyk}{1,0,0,0}
 \definecolor{MAGENTA}{cmyk}{0,1,0,0}
 \definecolor{YELLOW}{cmyk}{0,0,1,0}
}

\usepackage{xcolor}
\usepackage{soul}
\newcommand{\ket}[1]{\ensuremath{\left|#1\right\rangle}}

\definecolor{blue}{rgb}{0,0,1}
\definecolor{red}{rgb}{1,0,0}
\definecolor{green}{rgb}{0,1,0}

\usepackage{soul}

\makeatother

\begin{document}
\title{Demonstrating advantages of dynamic quantum circuits \\ on a hybrid superconducting qubit--cavity processor}

\author{Hongbo Wu}
\thanks{These authors contributed equally to this work.}
\affiliation{Southern University of Science and Technology, Shenzhen 518055, China}
\affiliation{International Quantum Academy, Shenzhen 518048, China.}

\author{Ling Hu}
\email{huling@iqasz.cn}
\thanks{These authors contributed equally to this work.}
\affiliation{International Quantum Academy, Shenzhen 518048, China.}
\affiliation{Shenzhen Branch, Hefei National Laboratory, Shenzhen 518048, China.}

\author{Jiasheng Mai}
\thanks{These authors contributed equally to this work.}
\affiliation{Southern University of Science and Technology, Shenzhen 518055, China}
\affiliation{International Quantum Academy, Shenzhen 518048, China.}

\author{Munan Zhang}
\thanks{These authors contributed equally to this work.}
\affiliation{School of Data Science, The Chinese University of Hong Kong, Shenzhen,
Guangdong, 518172, China}

\author{Libo Zhang}
\affiliation{Southern University of Science and Technology, Shenzhen 518055, China}
\affiliation{International Quantum Academy, Shenzhen 518048, China.}

\author{Yanyan Cai}
\affiliation{Southern University of Science and Technology, Shenzhen 518055, China}
\affiliation{International Quantum Academy, Shenzhen 518048, China.}

\author{Xiaowei Deng}
\affiliation{International Quantum Academy, Shenzhen 518048, China.}

\author{Pan Zheng}
\affiliation{International Quantum Academy, Shenzhen 518048, China.}

\author{Zhongchu Ni}
\affiliation{International Quantum Academy, Shenzhen 518048, China.}

\author{Song Liu}
\affiliation{International Quantum Academy, Shenzhen 518048, China.}
\affiliation{Shenzhen Branch, Hefei National Laboratory, Shenzhen 518048, China.}

\author{Kun Fang}
\email{kunfang@cuhk.edu.cn}
\affiliation{School of Data Science, The Chinese University of Hong Kong, Shenzhen,
Guangdong, 518172, China}

\author{Dapeng Yu}
\affiliation{International Quantum Academy, Shenzhen 518048, China.}
\affiliation{Shenzhen Branch, Hefei National Laboratory, Shenzhen 518048, China.}

\author{Yuan Xu}
\email{xuyuan@iqasz.cn}
\affiliation{International Quantum Academy, Shenzhen 518048, China.}
\affiliation{Shenzhen Branch, Hefei National Laboratory, Shenzhen 518048, China.}


\begin{abstract}
\textbf{Dynamic quantum circuits (DQCs) provide a hardware-efficient route to quantum computing by reducing physical-qubit overhead and compressing circuit topology through mid-circuit measurements, qubit reset and reuse, and classical feed-forward control. Here, we demonstrate the advantages of DQCs on a single hybrid superconducting qubit--cavity processor by implementing a hierarchy of algorithms with increasing complexity. This hybrid architecture consists of a high-dimensional cavity qudit serving as the computational register and a dispersively coupled superconducting transmon ancilla that is repeatedly measured, reset, and reused to enable dynamic control.
Using this device, we implement a $10$-bit Bernstein--Vazirani algorithm with an average success probability of $82\%$, surpassing state-of-the-art dynamic and static implementations in both scale and performance; an $8$-bit quantum phase-estimation protocol with estimation errors below $10^{-3}$; and the first dynamic-circuit implementation of Shor's algorithm on a superconducting platform, factoring $15$ over all coprime bases with squared statistical overlap values above $99.8\%$. 
These results provide concrete benchmarks for future DQC implementations and highlight the versatile advantages of DQCs with the hybrid qubit--qudit architecture, establishing it as a promising route toward scalable, programmable quantum computation.}
\end{abstract}
\maketitle
\vskip 0.5cm

\maketitle

\section{Introduction}

Building a scalable, programmable quantum computer capable of solving classically intractable problems~\cite{nielsen2010quantum}, such as integer factorization~\cite{shor1994algorithms}, quantum simulation~\cite{lloyd1996universal}, and quantum machine learning~\cite{biamonte2017quantum}, stands as one of the most ambitious goals in modern quantum information science. Scalability demands hardware-efficient architectures that minimize physical resource overhead, while programmability requires the flexibility to execute diverse quantum algorithms on a single physical platform~\cite{debnath2016demonstration}. A major bottleneck in meeting these requirements is the mismatch between a quantum processor's native connectivity and the interaction graph of the target circuit. Addressing this mismatch typically requires either hardware redesign with more suitable connectivity or the insertion of SWAP gates~\cite{siraichi2019qubit,yu2024symmetry}, leading to substantial architectural complexity and increased circuit depth~\cite{awschalom2025challenges}. Furthermore, compiling arbitrary algorithms onto fixed hardware topologies becomes increasingly challenging as system size grows, hindering both scalability and programmability in practical implementations. 

\begin{figure}[tb]
\centering
\includegraphics{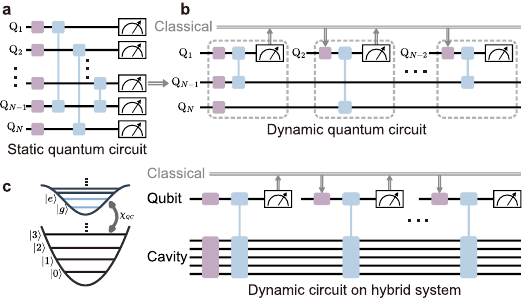}
\caption{\textbf{Schematic illustration of DQCs on a hybrid superconducting qubit--cavity processor.} 
\textbf{a} Conventional static quantum circuit with $N$ qubits. 
\textbf{b} Dynamic circuits: the upper qubit is measured, reset, and reused sequentially with real-time feed-forward. 
\textbf{c} Implementation of DQCs on a hybrid superconducting qubit--cavity system, where a transmon qubit is dispersively coupled to a long-lived cavity. The cavity's multi-level Hilbert space serves as the computational register and the transmon is repeatedly measured and reused.}
\label{fig: dynamic illustration}
\end{figure}

Dynamic quantum circuits (DQCs) offer a promising pathway to mitigate these challenges by leveraging mid-circuit measurements, qubit reset and reuse, and classical feed-forward control~\cite{corcoles2021exploiting}. Unlike conventional static circuits, where measurements only occur at the end (Fig.~\ref{fig: dynamic illustration}a), DQCs enable adaptive control and qubit reuse through mid-circuit measurements, significantly reducing physical-qubit overhead (Fig.~\ref{fig: dynamic illustration}b) and shortening effective circuit depth~\cite{piroli2021quantum,lu2022measurement,smith2024constant,baumer2024efficient}, thereby enhancing scalability~\cite{hua2023caqr,fang2026dynamic}. Consequently, DQCs have become central to emerging algorithmic designs~\cite{foss-feig2023experimental, iqbal2024nonabelian} and error-correction architectures~\cite{ni2023beating, gupta2024encoding, acharya2025quantum, bluvstein2026faulttolerant}.

Despite these conceptual advantages, DQCs present significant experimental challenges, demanding fast, high-fidelity, quantum non-demolition (QND) mid-circuit measurements, rapid resets, and low-latency classical signal processing, all of which impose stringent constraints on current quantum hardware. While dynamic circuits have recently been demonstrated on several platforms~\cite{corcoles2021exploiting,monz2016realization, martin2012experimental,hua2023caqr,baumer2024quantum,bluvstein2026faulttolerant}, most experiments have remained largely confined to individual tasks. A more rigorous and meaningful benchmark is to execute a sequence of diverse algorithms with increasing complexity on the same programmable processor, while maintaining a consistent performance advantage over static-circuit implementations.

In this work, we demonstrate, for the first time, the advantages of dynamic circuits across multiple quantum algorithms on a single hybrid superconducting qubit--cavity processor (Fig.~\ref{fig: dynamic illustration}c). This processor consists of a high-dimensional qudit encoded in a high-quality superconducting microwave cavity serving as the computational register, and a dispersively coupled superconducting transmon ancilla that is repeatedly measured, reset, and reused to enable dynamic control. This hybrid architecture offers intrinsic hardware efficiency and is naturally compatible with dynamic-circuit implementations for diverse algorithms~\cite{Crane2024,ma2025,liu2026}. Experimentally, we demonstrate the advantages of DQCs through a hierarchy of algorithms with increasing complexity: from the Bernstein--Vazirani (BV) algorithm~\cite{bernstein1993quantum} to quantum phase estimation (QPE)~\cite{kitaev1995quantum, nielsen2010quantum} and finally to Shor's algorithm~\cite{shor1994algorithms}. Specifically, we first realize a $10$-bit BV algorithm with an average success probability of $82\%$, surpassing state-of-the-art dynamic~\cite{hua2023caqr} and static~\cite{wright2019benchmarking} implementations in both scale and performance. We then implement an 8-bit QPE protocol, with estimation errors for all target phases remaining below $10^{-3}$, nearly an order of magnitude lower than previous results~\cite{corcoles2021exploiting}. Finally, we realize the first dynamic-circuit implementation of Shor's algorithm on a superconducting platform for factoring $15$, obtaining squared statistical overlap (SSO) values above $99.8\%$ for all six coprime bases, with errors suppressed by almost 10 times over previous implementations~\cite{monz2016realization}. These benchmarks span canonical milestones in quantum computing, from early demonstrations of quantum advantage to algorithms of practical significance, thereby highlighting the hybrid qubit--qudit paradigm with DQCs as a viable, hardware-efficient route toward scalable, programmable quantum computation.

\section{Results}

\subsection{Hybrid quantum processor}

The hybrid superconducting qubit--cavity processor for implementing the DQCs is constructed using a three-dimensional (3D) circuit quantum electrodynamics (cQED) device~\cite{blais2021,joshi2021,cai2021}. This device comprises a 3D coaxial superconducting microwave cavity~\cite{reagor2016quantum}, a superconducting transmon ancilla, and a Purcell-filtered stripline readout resonator~\cite{axline2016}. The 3D cavity, with coherence times $T_{1}^\mathrm{C}=0.64~\mathrm{ms}$ and $T_{2}^\mathrm{C}=1.01~\mathrm{ms}$, encodes the high-dimensional qudit that serves as the computational register in dynamic circuits. Its long coherence time enables coherent preservation of algorithmic states in DQCs with an efficient Fock-space encoding. The transmon ancilla, with coherence times $T_{1}^\mathrm{Q}=91~\mu\mathrm{s}$ and $T_{2}^\mathrm{Q}=124~\mu\mathrm{s}$, is dispersively coupled to the cavity with interaction strength $\chi_\mathrm{QC}/2\pi = 2.59~\mathrm{MHz}$, allowing efficient universal control of the hybrid qubit--qudit system~\cite{Krastanov2015Universal,heeres2017,eickbusch2022,diringer2024,huang2026}. Through its dispersive coupling to the readout resonator, the transmon ancilla is repeatedly measured and rapidly reset, facilitating dynamic circuit operations and real-time feed-forward control. Detailed device parameters are provided in Supplementary Sec.~I.

\subsection{Bernstein--Vazirani algorithm}
\begin{figure*}
\includegraphics{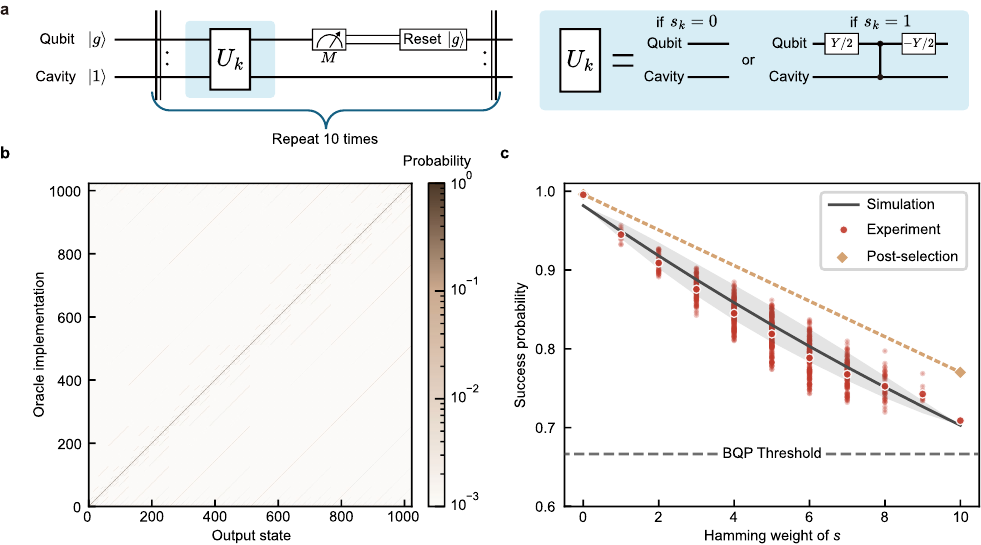} 
\caption{\textbf{Experimental realization of DQC for the 10-bit BV algorithm.}
\textbf{a} DQC sequence for the BV algorithm. The protocol comprises 10 iterative cycles, each corresponding to one bit $s_k$ of the hidden string $s$. The system is initialized in $\lvert g,1\rangle$. Each cycle contains a conditional operation $U_k$, a mid-circuit measurement ($M$), and a measurement-based qubit reset. The operation $U_k$ is conditioned on $s_k$: applying a CNOT-type operation if $s_k=1$, while idling if $s_k=0$.
\textbf{b} Measured output-state probability distribution for all $2^{10}=1024$ oracle configurations, each obtained from $45{,}000$ repetitions. The correct outputs correspond to the dominant diagonal elements. 
\textbf{c} Algorithmic success probability versus the Hamming weight (HW) of the hidden string \(s\). Red dots and the black curve denote the experimental and simulated success probabilities, respectively. The success probability ranges from 70.9\% ($\mathrm{HW}=10$) to 99.6\% ($\mathrm{HW}=0$), with an overall average of 82.0\%. Post-selecting the cavity in \(|1\rangle\) improves the success probability (orange dashed line) by about 6\% for $\mathrm{HW}=10$ and negligibly for $\mathrm{HW}=0$. All experimental instances exceed the standard \(2/3\) BQP threshold.}
\label{fig2} 
\end{figure*}

Using this hybrid processor, we first perform a dynamic-circuit implementation of the BV algorithm~\cite{bernstein1993quantum}, one of the earliest examples of provable quantum advantage and a standard benchmark for quantum processors. The BV algorithm determines a hidden bit string $s$ encoded in the oracle function $f(x)=s\cdot x \; (\mathrm{mod}\;2)$. Classically, this requires $n$ oracle queries to identify an $n$-bit string, whereas the quantum BV algorithm needs only a single oracle call. Although the quantum BV algorithm has been demonstrated on various platforms~\cite{hua2023caqr, fallek2016transport, debnath2016demonstration, linke2017experimental, wright2019benchmarking, roy2020programmable, blinov2021comparison, pokharel2023demonstration, mundada2023experimental, reichardt2025faulttolerant}, most implementions have operated within the static-circuit paradigm.

A fully dynamic compilation of any $n$-bit BV algorithm requires only $2$ physical qubits~\cite{decross2023qubit,fang2026dynamic}. In our experiment, we implement the BV algorithm via dynamic circuits for a $10$-bit string $s$, corresponding to $2^{10}=1024$ possible oracle functions. As shown in Fig.~\ref{fig2}a, the $k$-th bit $s_k$ determines the application of the oracle function $U_k$, i.e., applying a CNOT-type operation when $s_k=1$ and idling when $s_k=0$. The input state is fixed to the ground state $|g\rangle$ for the qubit and Fock state $|1\rangle$ for the cavity, and each circuit is repeated $45{,}000$ times. 

Figure~\ref{fig2}b shows the output state probability distributions for all $1024$ oracle realizations, where the correct output state always appears as the dominant diagonal element. We then calculate the success probability as the SSO $\left( \sum_i{ \sqrt{p_i^{\mathrm{meas}} p_i^{\mathrm{ideal}}} } \right)^2$~\cite{chiaverini2005implementation,monz2016realization} between the measured and ideal state probability distributions. The experimental results, plotted as a function of the Hamming weight of $s$ in Fig.~\ref{fig2}c, yield an average SSO of $82\%$ for all oracle realizations, with a minimum of $70.9\%$ for $s=1111111111$ and a maximum of $99.6\%$ for $s=0000000000$. Notably, all oracle implementations exceed the standard $2/3$ success probability threshold in the bounded-error quantum polynomial-time (BQP) computation~\cite{bernstein1993quantum}. Compared with previous benchmarks, our implementation achieves both higher success probability and larger problem size. It surpasses the 5-bit dynamic BV implementation in Ref.~\cite{hua2023caqr} (maximum success probability $\sim64\%$) and the 10-bit static implementation in Ref.~\cite{wright2019benchmarking} (average success probability $\sim78\%$). Post-selecting the cavity in Fock state $\lvert1\rangle$ at the end of the circuit improves the overlap for $s = 1111111111$ by about $6\%$. Numerical simulations identify qubit and cavity decay as the dominant error sources in the experiment (see Supplementary Sec.~III-A).

\subsection{Quantum phase estimation}

\begin{figure*}
\includegraphics{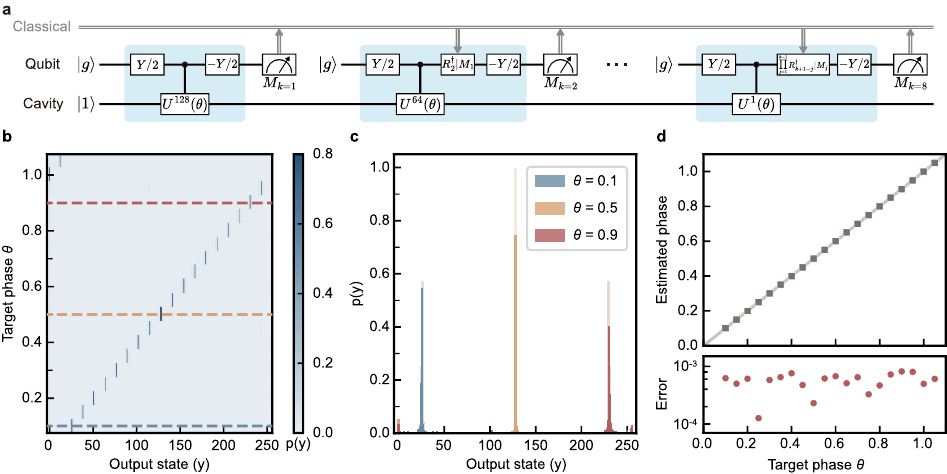} 
\caption{\textbf{Experimental realization of DQC for the 8-bit QPE algorithm.} 
\textbf{a} DQC sequence for the QPE algorithm. The cavity is initialized in Fock state $|1\rangle$, an eigenstate of the unitary operator $U = e^{i2\pi\theta a^\dagger a}$, while the transmon is initialized in ground state $|g\rangle$ and repeatedly measured and reset. The protocol consists of eight adaptive cycles, each including controlled-$U^{2^{8-k}}$ operations, single-qubit rotations, a mid-circuit measurement $M_k$, and classically conditioned phase gates implemented via virtual phase rotations. 
\textbf{b} Measured probability distributions of the binary output states $y$ for target phases $\theta \in \{0.1, 0.15, \ldots, 1.05\}$, obtained from $45{,}000$ repetitions of the experiment for each phase.
\textbf{c} Three representative horizontal cuts from \textbf{b}, showing the measured (solid bars) and ideal (shaded bars) probability distributions $p(y)$ for target phases $\theta = 0.1,\ 0.5,$ and $0.9$. 
\textbf{d} Estimated phase $\tilde{\theta}$ (top panel) as a function of the target phase $\theta$, showing good agreement with the ideal prediction (solid line), with the estimation errors shown in the bottom panel. For each phase, the total measurement resources are fixed at $R = 200$, and the circuit containing $m = 8$ measurement rounds is repeated $\lfloor R/m \rfloor$ times. The full dataset of $45{,}000$ circuit repetitions is divided into $1800$ independent phase-estimation trials to compute the error bars, which are smaller than the symbol size and thus are not shown. }
\label{fig3} 
\end{figure*}

Beyond the BV algorithm, we further demonstrate the advantage of dynamic circuits using the QPE algorithm, which requires more complex real-time feed-forward control. The QPE algorithm estimates the eigenphase of a unitary operator~\cite{kitaev1995quantum} and serves as a key subroutine in algorithms such as Shor's algorithm~\cite{shor1994algorithms}, quantum counting algorithm~\cite{brassard1998quantum}, and linear equation solving algorithm~\cite{harrow2009quantum}. Here, we consider the unitary operator $U(\theta)=e^{i2\pi\theta a^\dagger a}$ acting on the cavity qudit, where $a^\dagger$ and $a$ are the cavity ladder operators, and the Fock state $|n\rangle$ satisfies $a^\dagger a |n\rangle=n|n\rangle$. We use the single-photon Fock state $|1\rangle$ as the input eigenstate, for which $U(\theta)|1\rangle=e^{i2\pi\theta}|1\rangle$. In the standard static formulation of QPE, estimating the phase $\theta$ to $p$ bits of precision requires approximately $p$ ancilla qubits~\cite{nielsen2010quantum}, whereas a fully dynamic compilation requires only a single reusable ancilla qubit~\cite{dobsicek2007arbitrary}.

The dynamic QPE circuit for estimating $\theta\in[0.1,1.1)$ with $8$-bit precision is shown in Fig~\ref{fig3}a, which requires $m=8$ rounds of adaptive estimation steps, which are realized with a Ramsey-type sequence. For the $k$-th step ($k\ge 2$), a sequence of conditional phase gates $\prod_{j=1}^{k-1} R_{k+1-j}^\dagger \mid M_j = \text{diag}(1, e^{-\sum_{j=1}^{k-1} 2\pi i M_j/2^{k+1-j}})$ is applied on the ancilla, where the conditional phase depends on all preceding measurement outcomes $M_j\in\{0,1\}$ with $j=1,2...k-1$. In our experiment, we implement the sequence of conditional phase gates via a single virtual phase rotation~\cite{mckay2017} by combining all accumulated phases based on the measurement outcomes, effectively reducing the circuit depth.

Figure~\ref{fig3}b shows the measured output-state probability distributions as a function of the target phase $\theta$, obtained from $m=8$ mid-circuit measurements. For each target phase, the circuit is repeated $45{,}000$ times to reconstruct the output distribution. Three representative horizontal cuts at $\theta=0.1$, $0.5$, and $0.9$ are presented in Fig.~\ref{fig3}c, showing good agreement with theoretical predictions.
To enable a fair comparison with previous experimental QPE implementations, we adopt the same resource definition as Ref.~\cite{corcoles2021exploiting}, where a single qubit measurement is counted as one unit of resource. The total measurement resources for each phase estimation are fixed at $R=200$, the circuit is therefore repeated $\lfloor R/m \rfloor = 25$ times. The full dataset of $45{,}000$ circuit repetitions is thus divided into $1800$ independent phase-estimation trials. The estimated phase $\tilde{\theta}$ is extracted by taking the weighted average of the two neighboring bitstrings with the largest combined probability~\cite{corcoles2021exploiting}.
Figure~\ref{fig3}d plots the estimated phases and the corresponding estimation errors as a function of the target phase. The estimation errors remain below $10^{-3}$ over the entire range, representing nearly an order-of-magnitude improvement over previous implementations~\cite{corcoles2021exploiting}. These results demonstrate high-precision phase estimation and highlight the advantages of DQCs for implementing the QPE algorithm with the hybrid system.

\subsection{Shor's algorithm}

\begin{figure*}
\includegraphics{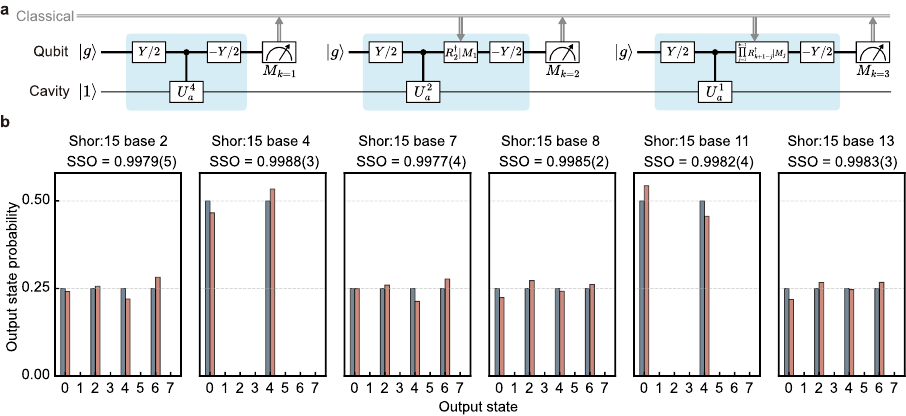} 
\caption{\textbf{Experimental realization of DQC for Shor's algorithm.}
\textbf{a} DQC sequence for the quantum order-finding subroutine. The cavity is initialized in the Fock state $|1\rangle$, while the qubit is initialized in $|g\rangle$ and repeatedly measured and reset. The protocol consists of sequential cycles implementing controlled modular exponentiation operations ($U_a^{4}$, $U_a^{2}$, and $U_a^{1}$), interleaved with single-qubit rotations and mid-circuit measurements $M_k$. The first measurement yields a deterministic outcome $|g\rangle$, and the experiment proceeds from the subsequent cycles. Classically conditioned phase rotations are applied based on prior measurement outcomes.
\textbf{b} Measured (red bars) and ideal (blue bars) state probability distributions for coprime bases $a \in \{2, 4, 7, 8, 11, 13\}$. The SSO between each measured and ideal distribution is calculated and indicated above each panel. }
\label{fig4} 
\end{figure*}

Building on these demonstrations, we further implement Shor's algorithm, which factors integers exponentially faster than the best-known classical algorithms and remains one of the most compelling examples of quantum advantage with practical relevance~\cite{shor1994algorithms}. Experimental demonstrations of Shor's algorithm have been reported across various physical platforms, including nuclear magnetic resonance~\cite{vandersypen2001experimental}, photonic systems~\cite{lanyon2007experimental, lu2007demonstration, politi2009shor, martin2012experimental}, trapped ions~\cite{monz2016realization}, and superconducting systems~\cite{lucero2012computing}. However, most of these implementations rely on conventional static circuits with substantial resource overhead and are often restricted to selected computational branches rather than implementing the full algorithm with the flexibility to accommodate arbitrary coprime bases. A faithful realization should preserve this algorithmic generality, because the coprime base is chosen at random and directly determines the overall success probability.

By extending the adaptive feed-forward control used in the dynamic QPE circuit and exploiting the cavity's high-dimensional Fock space as the computational register, we achieve a faithful implementation of Shor's algorithm for factoring $N=15$ over all coprime bases $a\in\{2,4,7,8,11,13\}$~\footnote{Note that when \(a = N-1\), we always have \(a^2 \equiv 1 \pmod{N}\). Hence, \(a\) has order \(2\), and no order-finding algorithm is required. Consequently, the case \(a=14\) is excluded here.}, with the experimental circuit shown in Fig.~\ref{fig4}a. The Fock states of the cavity qudit directly encode the basis states required for the modular-exponentiation operation $U_a$. The protocol retains the iterative phase-estimation structure and requires three adaptive measurement rounds. Each round involves a controlled modular-exponentiation operation and a classically controlled virtual phase rotation on the ancilla, with the phase determined by preceding measurement outcomes. Since $U_a^4 = I$ for this specific factoring problem, we omit the first measurement round and implement the controlled operations in subsequent cycles using quantum numerical optimal control pulses~\cite{khaneja2005,heeres2017}.

Figure~\ref{fig4}b shows the experimental results, where the three measurement outcomes are converted from binary strings to decimal values to determine the period. The performance is evaluated using the SSO, yielding values of \{0.9979(5), 0.9988(3), 0.9977(4), 0.9985(2), 0.9982(4), 0.9983(3)\} for the six coprime bases, respectively. These results significantly exceed the state-of-the-art in previous dynamic-circuit implementations on an ion-trap system~\cite{monz2016realization}, which achieved SSO values only in the range of $0.90-0.97$. Our demonstration constitutes the first high-performance dynamic-circuit implementation of Shor's algorithm on a superconducting platform,  highlighting the advantages of the long-lived cavity qudit and the hybrid qudit--qubit architecture for dynamic quantum computation. Numerical simulations indicate that the remaining errors are dominated by imperfections in the numerical pulses at higher Fock states and by qubit decay during the feed-forward process (see Supplementary Sec. III-C).

\maketitle
\section{Conclusion}
In conclusion, we have experimentally demonstrated the consistent advantages of DQCs across multiple quantum algorithms on a single hybrid superconducting qubit--cavity processor. By leveraging high-fidelity mid-circuit measurements, efficient qubit reset and reuse, fast classical feed-forward control, and hardware-efficient high-dimensional cavity encoding, we executed a hierarchy of three increasingly complex quantum algorithms---from the BV algorithm to QPE and Shor's algorithm---on the same programmable processor. In each case, our dynamic-circuit implementation substantially reduced physical-qubit overhead and achieved superior performance relative to both conventional static and prior dynamic implementations~\cite{hua2023caqr,wright2019benchmarking,corcoles2021exploiting,monz2016realization}, providing concrete benchmarks and highlighting architectural advantages of the hybrid paradigm with DQCs. 

These results establish the hybrid superconducting qubit--oscillator architecture~\cite{liu2026} as a hardware-efficient, programmable platform for dynamic quantum computation. Scaling this architecture to multiple interconnected cavities and qubits, while further improving coherence times~\cite{milul2023,bland2025}, would enable versatile applications in quantum error correction~\cite{cai2021,ni2023beating}, quantum metrology~\cite{fadel2024}, and quantum chemistry and gauge theory simulations~\cite{Crane2024,dutta2024}. Furthermore, the dynamic-circuit technique demonstrated here could be readily extended to other hybrid qubit-oscillator systems, including trapped ions or neutral atoms coupled to bosonic motional modes~\cite{hou2024,bazavan2026}, and superconducting qubits coupled to acoustic-wave phonons~\cite{chu2017,satzinger2018} or ferromagnetic magnons~\cite{Yutaka2015,xuda2023}, opening promising avenues for dynamic quantum computation on diverse physical platforms.

\smallskip{}


%

\clearpage{}

\noindent \textbf{\large{}Data availability}{\large\par}

\noindent All data generated or analysed during this study are available
within the paper and its Supplementary Information. Further source
data will be made available on reasonable request.

\smallskip{}

\noindent \textbf{\large{}Code availability}{\large\par}

\noindent The code used to solve the equations presented in the Supplementary
Information will be made available on reasonable request.

\smallskip{}

\noindent \textbf{\large{}Acknowledgment}{\large\par}

\noindent This work was supported by the National Natural Science Foundation of China (Grants No.~12422416, No.~12274198, No.~12374471), the Innovation Program for Quantum Science and Technology (Grants No.~2024ZD0302300, No.~2021ZD0301703), the Guangdong Basic and Applied Basic Research Foundation (Grant No.~2024B1515020013), and the Shenzhen Science and Technology Program (Grants No.~RCYX20221008092907026 and No. QNXMA20250701092312018). K.F. and M.Z. are supported in part by the National Natural Science Foundation of China (Grants No. 92470113 and No. 12404569), the Shenzhen Science and Technology Program (Grants No. QNXMB20250701091826036 and No. JCYJ20240813113519025), the Shenzhen International Quantum Academy (Grant No. SIQA2025KFKT03). 

\smallskip{}

\noindent \textbf{\large{}Author contributions}{\large\par}

\noindent Y.X. and L.H. supervised the experiment. K.F. supervised the theoretical scheme. H.W. and J.M. performed the experiments, analyzed the experimental data and carried out numerical simulations under the supervision of L.H. and Y.X. L.Z. fabricated the superconducting qubit under the supervision of S.L. M.Z. provided theoretical analysis under the supervision of K.F. H.B., L.H., J.M., M.Z., K.F., and Y.X. wrote the manuscript with feedback from all authors. Y.X. and D.Y. supervised the project.

\smallskip{}

\noindent \textbf{\large{}Competing interests}{\large\par}

\noindent The authors declare no competing interests.

\smallskip{}

\noindent \textbf{\large{}Additional information}{\large\par}

\noindent \textbf{Supplementary information} The online version contains
supplementary material.

\noindent \textbf{Correspondence and requests for materials} should be addressed to Ling Hu, Kun Fang, or Yuan Xu.

\clearpage{}

\newpage   
\clearpage 
\onecolumngrid

\clearpage

\setcounter{section}{0}

\begin{center}
\title{Supplementary Information for \\ ``Demonstrating advantages of dynamic quantum circuits \\ on a hybrid superconducting qubit--cavity processor''}
\end{center}

\section{Experimental device and parameters}

\subsection{Experimental device and setup}
\begin{figure}[htbp]
	\includegraphics{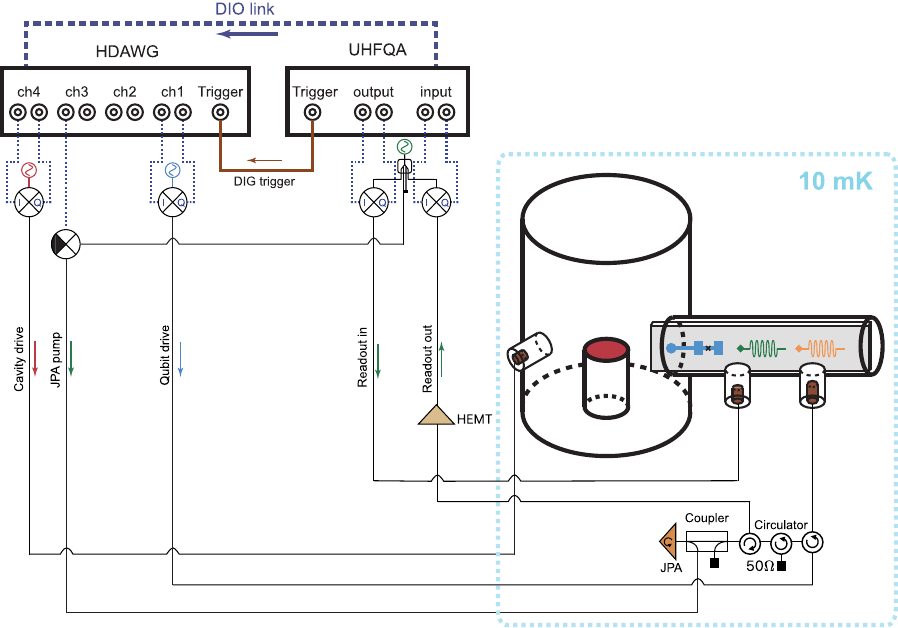}
	\caption{Measurement setup and experimental device schematic.}
	\label{figS1}
\end{figure}

Our experiments are conducted on a three-dimensional (3D) circuit quantum electrodynamics (cQED) architecture~\cite{blais2021}, similar to previous implementations~\cite{ni2023,deng2024quantum} (see schematic in Fig.~\ref{figS1}). The device comprises a high-purity (5N5) aluminum coaxial $\lambda/4$ cavity~\cite{reagor2016quantum}, a tantalum-film-based superconducting transmon ancilla~\cite{koch2007charge, Place2021}, and a Purcell-filtered stripline readout resonator~\cite{axline2016}. The Josephson junction, antenna pads of the transmon qubit, and the striplines of the readout resonator are lithographically defined on a single sapphire chip, which is inserted into a tunnel to couple to the coaxial cavity. To facilitate high-fidelity measurements while protecting the quantum system from environment-induced decoherence, the two-pad transmon ancilla couples to both the 3D cavity mode and the $\lambda/2$ stripline readout resonator. This readout resonator is cascaded with an additional $\lambda/2$ stripline that acts as a bandpass Purcell filter~\cite{axline2016}, enabling fast state interrogation while strongly suppressing environmental dissipation.

The experimental device is housed within a magnetic shield anchored to the base plate of a dilution refrigerator operating below $10~\mathrm{mK}$. Microwave control signals are synthesized using room-temperature electronics. Drive pulses for the ancilla and the cavity are generated via single-sideband in-phase and quadrature (IQ) modulation, with intermediate-frequency waveforms generated from an arbitrary waveform generator (Zurich Instruments, HDAWG). Similarly, the readout pulses are synthesized using a quantum analyzer (Zurich Instruments, UHFQA). These signals are routed to the device through coaxial cables equipped with a series of isolators, filters, and attenuators to suppress microwave reflections and thermal radiation. The output readout signal is processed through a multistage amplification chain, comprising a quantum-limited Josephson parametric amplifier (JPA) at the base temperature, a high-electron-mobility transistor (HEMT) amplifier at the $4~\mathrm{K}$ stage, and standard commercial low-noise amplifiers at room temperature. The amplified signal is downconverted via analog IQ demodulation that shares the same local oscillator as the readout generation, and subsequently digitized by the UHFQA. Precise timing synchronization between the HDAWG and UHFQA is maintained via digital triggers (DIG), while a direct DIO link between the UHFQA and HDAWG facilitates real-time feed-forward with a $700~\mathrm{ns}$ latency.

\subsection{System Hamiltonian and device parameters}

The 3D cQED system can be described by the dispersive Hamiltonian~\cite{blais2021}:
\begin{align}\label{H_sym}
	\begin{split}
		\hat{H}/\hbar
		 & =\omega_\mathrm{Q}\hat{a}_\mathrm{Q}^\dagger\hat{a}_\mathrm{Q}+\omega_\mathrm{C}\hat{a}_\mathrm{C}^\dagger\hat{a}_\mathrm{C}+\omega_\mathrm{R}\hat{a}_\mathrm{R}^\dagger\hat{a}_\mathrm{R}                                      \\
		 & -\left(\frac{K_\mathrm{Q}}{2}\hat{a}_\mathrm{Q}^{\dagger2}\hat{a}_\mathrm{Q}^2 + \frac{K_\mathrm{C}}{2}\hat{a}_\mathrm{C}^{\dagger2}\hat{a}_\mathrm{C}^2\right)                                       \\
		 & -\chi_{\mathrm{QC}}\hat{a}_{\mathrm{Q}}^{\dagger}\hat{a}_{\mathrm{Q}}\hat{a}_{\mathrm{C}}^{\dagger}\hat{a}_{\mathrm{C}}-\chi_{\mathrm{QR}}\hat{a}_{\mathrm{Q}}^{\dagger}\hat{a}_{\mathrm{Q}}\hat{a}_{\mathrm{R}}^{\dagger}\hat{a}_{\mathrm{R}},
	\end{split}
\end{align}
where $\omega_{\mathrm{Q,C,R}}$ denote the resonance frequencies of the ancilla qubit, the cavity, and the readout resonator, respectively. The operators $\hat{a}_{\mathrm{Q,C,R}}$ and $\hat{a}^\dagger_{\mathrm{Q,C,R}}$ represent their corresponding annihilation and creation operators. The terms $K_{\mathrm{Q,C}}$ denote the self-Kerr anharmonicities of the ancilla and the cavity mode, while $\chi_{\mathrm{QC}}$ and $\chi_{\mathrm{QR}}$ represent the dispersive cross-Kerr interactions between the ancilla and the respective linear modes. 

All these system parameters and coherence times, characterized via standard cQED protocols, are summarized in Table~\ref{tab:parameter_sum}. Additionally, higher-order dispersive terms not explicitly included in Eq.~(\ref{H_sym}), such as the second-order cross-Kerr interaction $\chi_\mathrm{QC}'\hat{a}_{\mathrm{Q}}\hat{a}_{\mathrm{Q}}^{\dagger}\hat{a}_{\mathrm{C}}^{\dagger2}\hat{a}_{\mathrm{C}}^{2}$ between the ancilla and the cavity, are also listed in the table. 

In our experiment, we characterize the coherence times of the three subsystems and list them in Table~\ref{tab:parameter_sum}. The transmon ancilla exhibits an energy relaxation time of $T_{1, \mathrm{Q}} \approx 91$~\textmu s and a Ramsey coherence time of $T_{2,\mathrm{Q}} \approx 124$~\textmu s, yielding a pure dephasing time of $T_{\phi,\mathrm{Q}} \approx 389$~\textmu s. For the cavity mode, we measure a single-photon lifetime of $T_{1, \mathrm{C}} \approx 641$~\textmu s and a Ramsey coherence time of $T_{2,\mathrm{C}} \approx 1013$~\textmu s, corresponding to a pure dephasing time of $T_{\phi, \mathrm{C}} \approx 4828$~\textmu s. 

\begin{table*}[htbp]
\renewcommand{\arraystretch}{1.3}
	\begin{ruledtabular}
		\centering
		\caption{Hamiltonian parameters and coherence times}
		\label{tab:parameter_sum}
		\begin{tabular}{cccccc}
			Measured parameter                                           & Cavity &   & Transmon qubit &   & Readout resonator \\
			\cline{1-6}
			Mode frequency $\omega_\mathrm{{C,Q,R}}/2\pi$                         & $6.532~\mathrm{GHz}$              &   & $4.952~\mathrm{GHz}$              &   & $8.564~\mathrm{GHz}$              \\
			Self-Kerr $K_\mathrm{{C,Q,R}}/2\pi$                                   & $9~\mathrm{kHz}$             &   & $215~\mathrm{MHz}$              &   & -            \\
			Cross-Kerr $\chi_\mathrm{{QC}}/2\pi,\chi_{\mathrm{QR}}/2\pi$                   &                & $2.59~\mathrm{MHz}$ &                & $2.1~\mathrm{MHz}$ &                \\
			Second-order Cross-Kerr $\chi_{\mathrm{QC}}^{'}/2\pi$ &                & $5.8~\mathrm{kHz}$ &                & - &                \\
			\cline{1-6}
			Relaxation time $T_1$                                     & 
            $641 $~\textmu s  
            &   & $91 $~\textmu s              &   & $55~\mathrm{ns}$              \\
            Decoherence time $T_2$                           & 
            $1013 $~\textmu s
            &   & $124$~\textmu s             &   & -              \\
			Pure dephasing time $T_\phi$                           & 
            $4828$~\textmu s
            &   & $389$~\textmu s             &   & -              \\
			Thermal population $P_{\mathrm{th}}$                                  & $0.4\%$              &   & $1.5\%$              &   & -              \\
		\end{tabular}
	\end{ruledtabular}
\end{table*}

\section{Experimental characterization}
\subsection{Measurement fidelity and active reset performance}
\begin{figure}[htbp]
\centering
\includegraphics{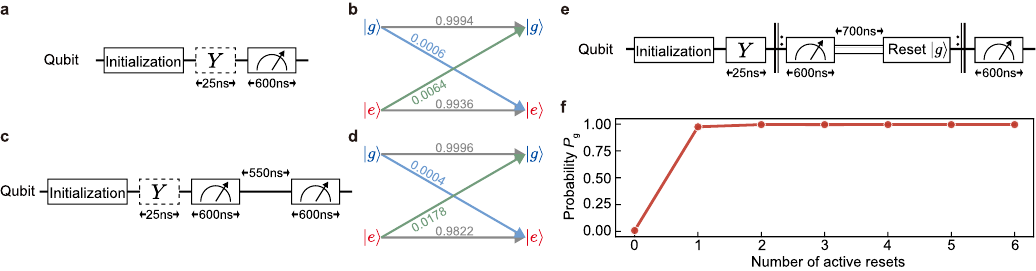}
    \caption{\textbf{Characterization of qubit readout and QND fidelity.} 
\textbf{a} Experimental sequence for evaluating the measurement fidelity. 
\textbf{b} Measured readout fidelities: $p_g = 0.9994$ and $p_e = 0.9936$. 
\textbf{c} Experimental sequence for extracting the QND fidelity.
\textbf{d} Measured QND fidelities: $p_{gg} = 0.9996$ and $p_{ee} = 0.9822$. 
\textbf{e} Experimental sequence for characterizing the active qubit reset with measurement-based feedback control. 
\textbf{f} Measured qubit ground-state probability as a function of the number of active resets.}
	\label{fig:QND_reset}
\end{figure}


Quantum algorithms compiled with dynamic quantum circuits (DQCs) in our experiment involve repeated implementations of mid-circuit measurements of the ancilla qubit. Thus, it is crucial to improve the qubit readout fidelity to enhance the overall performance. 

In our experiment, we employ a segmented readout pulse scheme~\cite{McClure2016Rapid}, with a total readout duration of $600\,\mathrm{ns}$, including a $100\,\mathrm{ns}$ high-power plateau. With the assistance of a Josephson parametric amplifier (JPA)~\cite{vijay2011observation}, we could achieve high-fidelity mid-circuit measurement of the ancilla qubit. Using the sequence shown in Fig.~\ref{fig:QND_reset}, we characterize the ancilla readout fidelity and its quantum nondemolition (QND) property~\cite{Hu2019BinomialQEC}. The system is initialized by post-selecting the ancilla in \(|g\rangle\) and the cavity in the even-parity subspace to remove residual thermal populations.

After the experimental characterization, we obtain an average measurement fidelity of $F = (P_g + P_e)/2 = 99.65\%$ and a QND fidelity of $F_{\mathrm{QND}} = (P_{g,g} + P_{e,e})/2 = 99.09\%$. Notably, the high ground-state readout fidelity $P_g = 99.94\%$ is essential for achieving a high computational fidelity of up to $99.6\%$ in the Bernstein--Vazirani (BV) algorithm for the trivial oracle $s = 0000000000$. The infidelity of the excited-state readout is predominantly limited by qubit energy relaxation.

Furthermore, we characterize the active reset sequence (Fig.~\ref{fig:QND_reset}e). After preparing the qubit in the excited state with a $25\,\mathrm{ns}$ $Y$ pulse, an initial measurement is performed. Conditioned on the real-time readout outcome, a feed-forward operation is applied to deterministically reset the qubit to $|g\rangle$. A subsequent verification measurement yields an active reset fidelity of $F_{\mathrm{reset}} = 97.48\%$. The remaining error is mainly limited by the readout infidelity and the ancilla decay during the $\sim 700\,\mathrm{ns}$ latency of the classical feed-forward.

\subsection{Controlled modular exponentiation operations in Shor's algorithm}
\label{sec:III-B}
In our experiment, the cavity serves as the computational register, where the Fock-state subspace spanned by $|0\rangle$ to $|14\rangle$ constitutes the computational space for factoring 15, with the register states represented in decimal form. Since \(U_a^{4}=I\) for all choices of \(a\), the first measurement cycle can be omitted, and the experiment effectively starts after the first measurement with the deterministic outcome \(m_{1}=0\). Both the controlled-\(U_a^{2}\) and controlled-\(U_a\) operations are implemented using numerical optimal control with the gradient ascent pulse engineering (GRAPE) method~\cite{khaneja2005,heeres2017}. To compensate for additional phases accumulated during the waiting and measurement processes due to Kerr and cross-Kerr interactions, we incorporate virtual phase corrections into the GRAPE pulse optimization.

Because the input state for \(CU_a^{2}\) is fixed as $|+,1\rangle$
where the second register is expressed in decimal notation, the target output state is constrained to
\begin{equation}
    CU_a^2 |+\rangle \otimes |1\rangle \rightarrow \frac{1}{\sqrt{2}} \left( |g\rangle \otimes |1\rangle + |e\rangle \otimes |a^2 \bmod{15}\rangle \right).
    \label{eq:state_transfer_cua2}
\end{equation}

Following the $CU_a^2$ operation, a subsequent Hadamard gate and measurement of the control qubit project the computational register onto one of two deterministic superposition states, depending on the measurement outcome $m_2 \in \{0,1\}$:
\begin{align}
    m_2 &= 0, \quad |\psi_0\rangle = \frac{1}{\sqrt{2}} \left( |1\rangle + |a^2 \bmod{15}\rangle \right), \label{eq:collapse_0} \\
    m_2 &= 1, \quad |\psi_1\rangle = \frac{1}{\sqrt{2}} \left( |1\rangle - |a^2 \bmod{15}\rangle \right). \label{eq:collapse_1}
\end{align}

This measurement-induced state collapse implies that the final modular multiplier $CU_a$ acts upon a conditionally prepared input. Accordingly, we formulate two parallel, feed-forward-dependent optimization targets:
\begin{equation}
    \begin{aligned}
        CU_a |+\rangle \otimes |\psi_0\rangle &\rightarrow \frac{1}{2}\Big( |g\rangle \otimes \left(|1\rangle + |a^2\bmod{15}\rangle\right) + |e\rangle \otimes \left(|a \bmod{15}\rangle + |a^3 \bmod{15}\rangle\right) \Big),
    \end{aligned}
    \label{eq:state_transfer_cua_0}
\end{equation}
\begin{equation}
    \begin{aligned}
        CU_a |+\rangle \otimes |\psi_1\rangle &\rightarrow \frac{1}{2}\Big( |g\rangle \otimes \left(|1\rangle - |a^2 \bmod{15}\rangle\right) 
        + |e\rangle \otimes \left(|a \bmod{15}\rangle - |a^3 \bmod{15}\rangle\right) \Big).
    \end{aligned}
    \label{eq:state_transfer_cua_1}
\end{equation}

\begin{figure*}[htbp] 
    \centering
    \includegraphics[width=\textwidth]{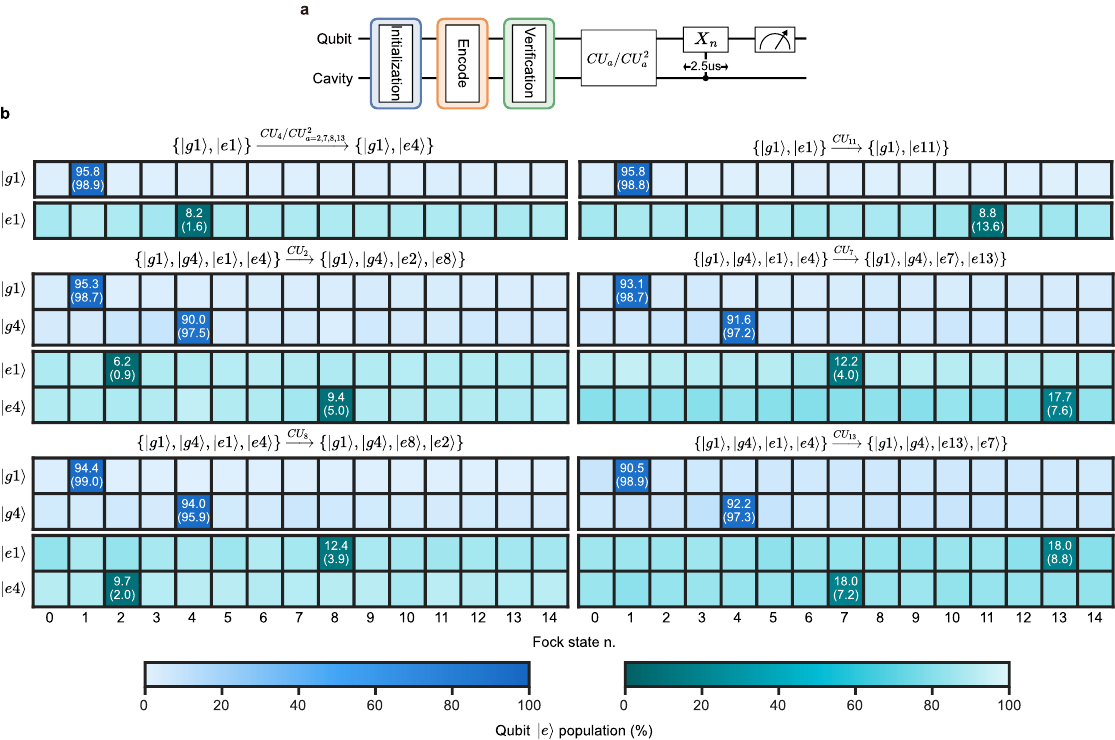} 
    \caption{\textbf{Characterizing the GRAPE-optimized controlled-unitary operations.}
\textbf{a} Experimental sequence for evaluating the state mapping performance. After initializing the system in $\ket{g,0}$, we encode the system in different input states, apply the optimized controlled-unitary operations, and measure the qubit excited-state populations following an $n$-photon-number-selective $\pi$ pulse.
\textbf{b} Measured qubit excited-state populations with the selective pulse applied on different photon numbers $n$ for the six controlled operations. The left labels denote the input states, and the target mappings are indicated above each matrix. The matrix elements represent the measured qubit excited-state populations after the photon-number-selective $\pi$ pulse, with the values in parentheses obtained from numerical simulations. }
\label{fig:grape_truth_table}
\end{figure*}

By strictly restricting the input states of the pulse-optimized unitary transformations to the actual physical states determined by measurement feed-forward, we drastically compress the parameter search space. This enables the synthesis of high-fidelity optimized control waveforms.

We experimentally characterize the optimized controlled unitary operations using the sequence shown in Fig.~\ref{fig:grape_truth_table}a. For factoring 15 in Shor's algorithm, there are six controlled unitary operations, corresponding to six distinct state mappings. In the experimental sequence, we first initialize the qubit--cavity system in different relevant states, then apply each controlled operation, and finally measure the qubit excited-state population after a photon-number-selective $\pi$ pulse on the ancilla qubit.  The measurement results, shown in Fig.~\ref{fig:grape_truth_table}b, clearly indicate that each unitary operation correctly maps arbitrary initial states within the physically relevant subspace to their target final states. 

When the ancilla is initialized in $|e\rangle$, a noticeable residual background population outside the target Fock state is observed. This background likely originates from the infidelity of the corresponding GRAPE operation, which fails to preserve the qubit state after the operation. This effect would become increasingly pronounced for mappings involving higher-photon-number Fock states.

To investigate this further, we perform numerical simulations of the process, with the simulation results shown in the parentheses in Fig.~\ref{fig:grape_truth_table}. In the simulation, similar residual backgrounds are observed when the ancilla is initialized in $|e\rangle$.  As the mappings involve higher Fock states and become more complex, the fidelity of the unitary operation gradually decreases, and the experiment-simulation discrepancy increases accordingly. This indicates that the high-photon-number regime is more sensitive to control imperfections, such as parameter mismatches, Hamiltonian parameter drifts, pulse distortions, and model inaccuracies, thereby limiting GRAPE control accuracy and achievable gate fidelity.

\section{Error analysis of dynamic quantum circuits}
To identify the dominant error sources and their contributions in our DQC experiments, we perform numerical simulations by solving the Lindblad master equation using QuTiP~\cite{johansson2013qutip}. In the simulation, we initialize the qubit--cavity system in the state $|g,0\rangle$ and include all decoherence processes, with the coherence times ($T_1$ and $T_{\phi}$) and thermal populations listed in Table~\ref{tab:parameter_sum}.


\subsection{Bernstein--Vazirani algorithm}
\begin{figure}[b]
\centering
\includegraphics{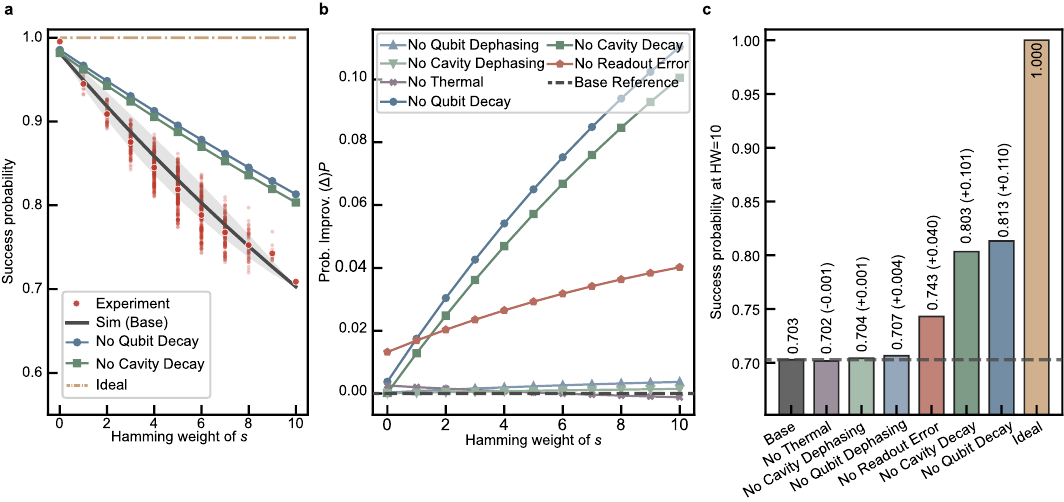}
	\caption{\textbf{Numerical simulation of the BV algorithm.}
    \textbf{a} Experimentally measured (red dots) and numerically simulated (solid lines) success probability of the BV algorithm as a function of the Hamming weight (HW) of the hidden bit string $s$. The base simulation (solid black line) includes all dissipative processes present in the experiment, and the ``No qubit decay'' and ``No cavity decay'' denote the simulations that exclude the qubit and cavity decay, respectively. The ideal case corresponds to the simulation with only ideal operations.
    \textbf{b} Success probability improvements as a function of the HW, obtained by comparing the simulated success probability with and without considering the specific error sources. \textbf{c} Simulated success probabilities with removing the specific error channels at $\mathrm{HW}=10$. The values of the bar heights are indicated, with the values in parentheses denoting the corresponding success probability improvement \(\Delta P\). Qubit and cavity decay are the dominant error sources, contributing infidelities of about 0.11 and 0.10, respectively. 
    }
	\label{fig:SM_BV}
\end{figure}

Figure~\ref{fig:SM_BV}(a) shows the simulated success probability of the BV algorithm as a function of the Hamming weight (HW) of the hidden bit string $s$. The base simulation (solid black line) includes all dissipative processes and shows good agreement with the experimental data (red markers) over the entire HW range, indicating that our theoretical noise model captures the dominant physical imperfections of the current hardware.

To isolate the contributions of individual error sources, we systematically deactivate specific dissipators in the simulation. Here, we consider six dissipative channels: qubit decay and dephasing, cavity decay and dephasing, qubit thermal excitation, and readout error. By comparing simulation results with and without a specific error channel, we can evaluate the error contribution via the increase in success probability, defined as $\Delta P = P_{\mathrm{modified}} - P_{\mathrm{base}}$. Figure~\ref{fig:SM_BV}(b) shows the success probability improvements for all six cases as a function of the HW, with the resulting success probability for HW =10 summarized in Fig.~\ref{fig:SM_BV}(c). 

Our simulation results reveal that the dominant error sources are qubit and cavity decay, which contribute infidelities of 0.11 and 0.10 to the success probability, respectively. The infidelity from the qubit decay primarily originates from the qubit energy relaxation during the feed-forward latency, accounting for 0.095 of the infidelity.  Pure dephasing of both the qubit and the cavity, as well as thermal excitations, have a negligible impact on the overall success probability ($< 1\%$).

\begin{figure}[t]
\centering
\includegraphics{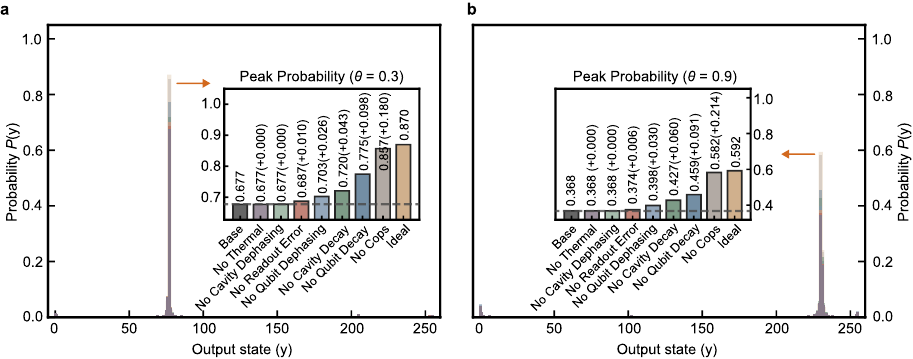}
	\caption{\textbf{Error analysis of the QPE experiment.}
\textbf{a} Measured and simulated probability distributions of the output state \(y\) for estimating the target phase \(\theta=0.3\). The inset shows the simulated peak probabilities with and without the specific dissipative channels to extract their contributions to the infidelity. The ``no cops'' result corresponds to the simulation without any collapse operators (retaining only readout infidelity) and agrees well with the ideal theoretical value. The base simulation includes all collapse channels and exhibits good agreement with the experimental peak probability.
\textbf{b} Error analysis similar to \textbf{a} but for $\theta = 0.9$. 
}
	\label{fig:PE}
\end{figure}

\begin{figure}[b]
\centering
\includegraphics{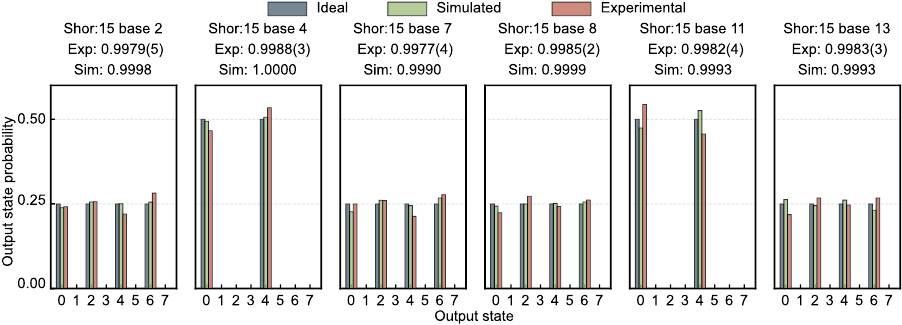}
\caption{\textbf{Measured and simulated probability distributions for Shor's algorithm with coprime bases $a \in \{2, 4, 7, 8, 11, 13\}$.}
In each panel, the experimental results (red bars) are compared with numerical simulations that include all identified error sources (green bars), and with the ideal theoretical distributions (blue bars). The corresponding SSO values are displayed above each panel.}
\label{fig:Shor}
\end{figure}

\subsection{Quantum phase estimation algorithm}
The dynamic QPE circuit for estimating a target phase $\theta$ to $m$-bit precision requires the implementation of $m$ controlled-unitary operations $\hat{U}^{2^{k-1}}(\theta)$ with $k = 1,2,\ldots,m$. In our hybrid qubit--cavity system, the operation $U(\theta)=e^{i2\pi\theta a^\dagger a}$ is realized by engineering a dispersive interaction duration of $\tau = 2\pi\theta/\chi$. Consequently, the operation $\hat{U}^{2^{k-1}}(\theta)$ can be naturally implemented by setting the dispersive interaction duration to $\tau_k = 2\pi\theta 2^{k-1}/\chi$. 

In our dynamic QPE experiment, we estimate the phase $\theta$ of the operator $U(\theta)$ to 8-bit precision using $m=8$ rounds of controlled-unitary operations. To identify the dominant error sources, we perform numerical simulations and compare the simulated results with and without specific error channels. The results are shown in the insets of Fig.~\ref{fig:PE}. Here, the ``no COPS'' case corresponds to the simulation without any decoherence channels, and the simulation result agrees well with the ideal theoretical case. The ``base'' case includes all collapse channels and exhibits good agreement with the experimental data, indicating that our numerical simulations accurately capture the experimental dynamics and validate the effectiveness of the error analysis. The numerical simulation results reveal that the phase estimation performance is mainly limited by the energy relaxation of both the cavity and the ancilla qubit.

\subsection{Shor's algorithm}

For the Shor's algorithm experiment, we also perform numerical simulations of the dynamic circuit with including all identified error sources. The simulation results, shown in Fig.~\ref{fig:Shor}, yield a squared statistical overlap (SSO) close to unity with respect to the ideal theoretical distribution. The remaining experimental errors likely originate from residual control non-idealities in the GRAPE-based manipulation of high-photon-number states. As discussed in Sec.~\ref{sec:III-B}, the fidelity of the unitary operations gradually degrades as the state mappings involve higher Fock states and become more complex.

\end{document}